\documentclass[twocolumn,prd,noshowpacs,nofootinbib,amsmath,amssymb,superscriptaddress]{revtex4-1}
\usepackage{graphicx,epsfig,psfrag,bm,amssymb}
\usepackage{dcolumn}
\usepackage{bm}
\usepackage{mciteplus}
\usepackage{color}
\usepackage{mathrsfs,amsmath, amsfonts,hepunits, color}
\usepackage{tikz}
\usepackage{slashed}
\usetikzlibrary{arrows,shapes}
\usetikzlibrary{trees}
\usetikzlibrary{matrix,arrows} 				
\usepackage{slashed}

\newcommand{\be}{\begin{eqnarray}}
\newcommand{\ee}{\end{eqnarray}}
\def\lsim{\mathrel{\rlap{\lower4pt\hbox{\hskip 0.5 pt$\sim$}}
    \raise1pt\hbox{$<$}}}                
\def\gsim{\mathrel{\rlap{\lower4pt\hbox{\hskip1pt$\sim$}}
    \raise1pt\hbox{$>$}}}

\newcommand{\s}{{\rm s}}

\def\lsim{\mathrel{\rlap{\lower4pt\hbox{\hskip1pt$\sim$}}
    \raise1pt\hbox{$<$}}}
\def\gsim{\mathrel{\rlap{\lower4pt\hbox{\hskip1pt$\sim$}}
    \raise1pt\hbox{$>$}}}

\newcommand{\apr}{A'}

\begin{document}

\title{ 
MeV-Scale Dark Matter Deep Underground
}
\author{ Eder Izaguirre}
\author{ Gordan Krnjaic}
 \affiliation{Perimeter Institute for Theoretical Physics, Waterloo, Ontario, Canada    }
\author{ Maxim Pospelov}
 \affiliation{Perimeter Institute for Theoretical Physics, Waterloo, Ontario, Canada    }
 \affiliation{Department of Physics and Astronomy, University of Victoria,  
Victoria, British Columbia, Canada}

\begin{abstract}

We demonstrate that current and planned underground neutrino experiments could offer a powerful probe of few-MeV dark matter when combined with a nearby high-intensity low-to-medium energy electron accelerator.
This experimental setup, an underground beam-dump experiment, is capable of decisively testing the thermal freeze-out mechanism for several natural  dark matter scenarios in this mass range. We present the sensitivity reach in terms of the mass-coupling parameter space of existing and planned detectors, such as Super-K, SNO+, and JUNO, in conjunction with a hypothetical 100 MeV energy accelerator. This setup can
also greatly extend the sensitivity of direct searches for new light weakly-coupled force-carriers
  independently of their connection to dark matter. 
\end{abstract}

\maketitle

%
%

\section{Introduction}

%

The existence of Dark Matter (DM) is clear evidence of physics 
beyond Standard Model (SM) and has inspired a major experimental effort to 
to uncover its particle nature. If DM achieves thermal equilibrium with the SM in the 
early universe, its present-day abundance can arise from DM annihilation with characteristic cross
section $\sigma v \sim 3\times 10^{-26} \cm^3/\s$. Alternatively if its 
abundance at late times is set by a primordial particle-antiparticle asymmetry, a thermal origin 
requires {\it at least} as large of an annihilation rate to 
avoid cosmological overproduction. For either scenario, this requirement sets a predictive target of opportunity to search for many of the simplest 
light DM models.

Current and planned direct and indirect detection, and collider experiments will cover a vast subset of DM masses
motivated by the thermal origin paradigm. However, significant gaps remain in our current search strategies for low-mass DM. Indeed, the  MeV-to-GeV DM mass window remains an elusive blind spot in the current search effort \cite{Cushman:2013zza}, despite the existence of viable models \cite{Boehm:2002yz, Boehm:2003hm, Pospelov:2007mp, Pospelov:2008zw, ArkaniHamed:2008qn, Pospelov:2008jd, Izaguirre:2015yja} -- including those invoked to explain the excess 511 keV photon line from the galactic bulge \cite{Jean:2003ci} with MeV scale DM annihilation into electron-positron pairs \cite{Boehm:2003hm, Pospelov:2007mp}. 
 Recent progress in our understanding of the status of MeV-scale DM has come from a combination of re-interpretation of surface-level proton-beam neutrino experiments results \cite{Batell:2009di, deNiverville:2011it, deNiverville:2012ij,deNiverville:2015mwa}, rare meson decays \cite{Bird:2004ts,Ablikim:2007ek,Adler:2004hp,Artamonov:2009sz,Dolan:2014ska}, electron beam dump experiments \cite{Izaguirre:2013uxa, Izaguirre:2014dua, Batell:2014mga, Izaguirre:2014bca}, B-factories \cite{Izaguirre:2013uxa, Essig:2013vha}, precision measurements \cite{Pospelov:2008zw, Izaguirre:2013uxa, Giudice:2012ms},  the CMB \cite{Finkbeiner:2011dx, Lin:2011gj,Galli:2011rz, Madhavacheril:2013cna, Ade:2013zuv},  and DM-electron scattering in direct detection experiments \cite{Essig:2012yx}. 

\begin{figure}[t!]

 \vspace{1cm}
\includegraphics[width=8.9cm]{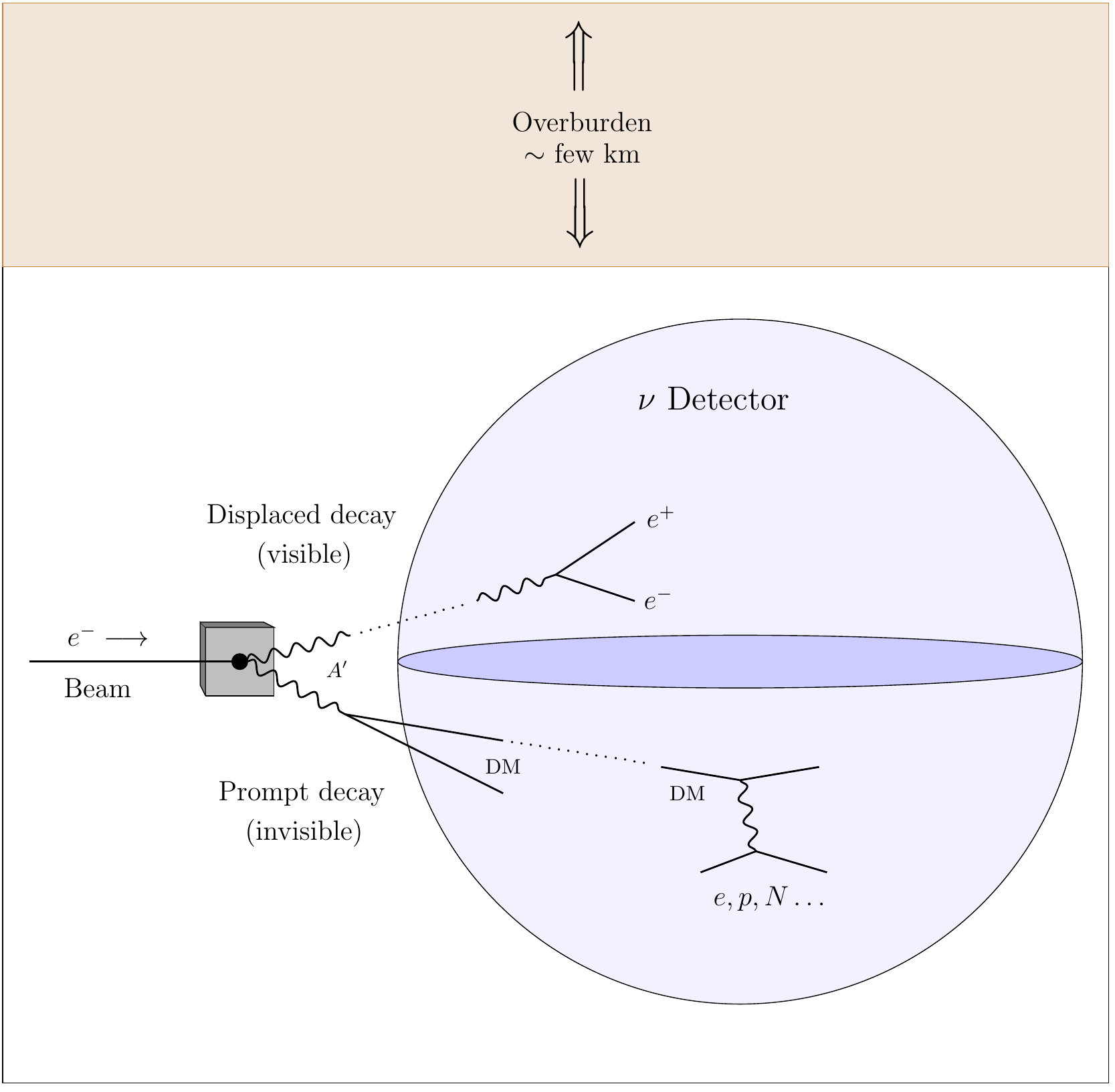}
  \caption{
  Schematic diagram of the proposed setup: a high intensity electron accelerator is placed in the vicinity of a large, underground neutrino detector. 
  The electron beam impinges on a fixed-target or beam-dump to produce a dark force-carrier $\apr$, which can decay either visibly to $e^+e^-$ or 
  to DM particles. If the $\apr$ decays visibly and is long lived, it can enter the detector and directly deposit a large electromagnetic 
  signal. If the $\apr$ decays invisibly to DM, the daughter particles inherit forward-peaked kinematics and scatter in the detector inducing observable target-particle
  recoils. 
  }
\label{fig:schematic}
\end{figure}


\begin{figure*}[t!]
 \vspace{0.cm}
 \hspace{-0.3cm}
\includegraphics[width=8.6cm]{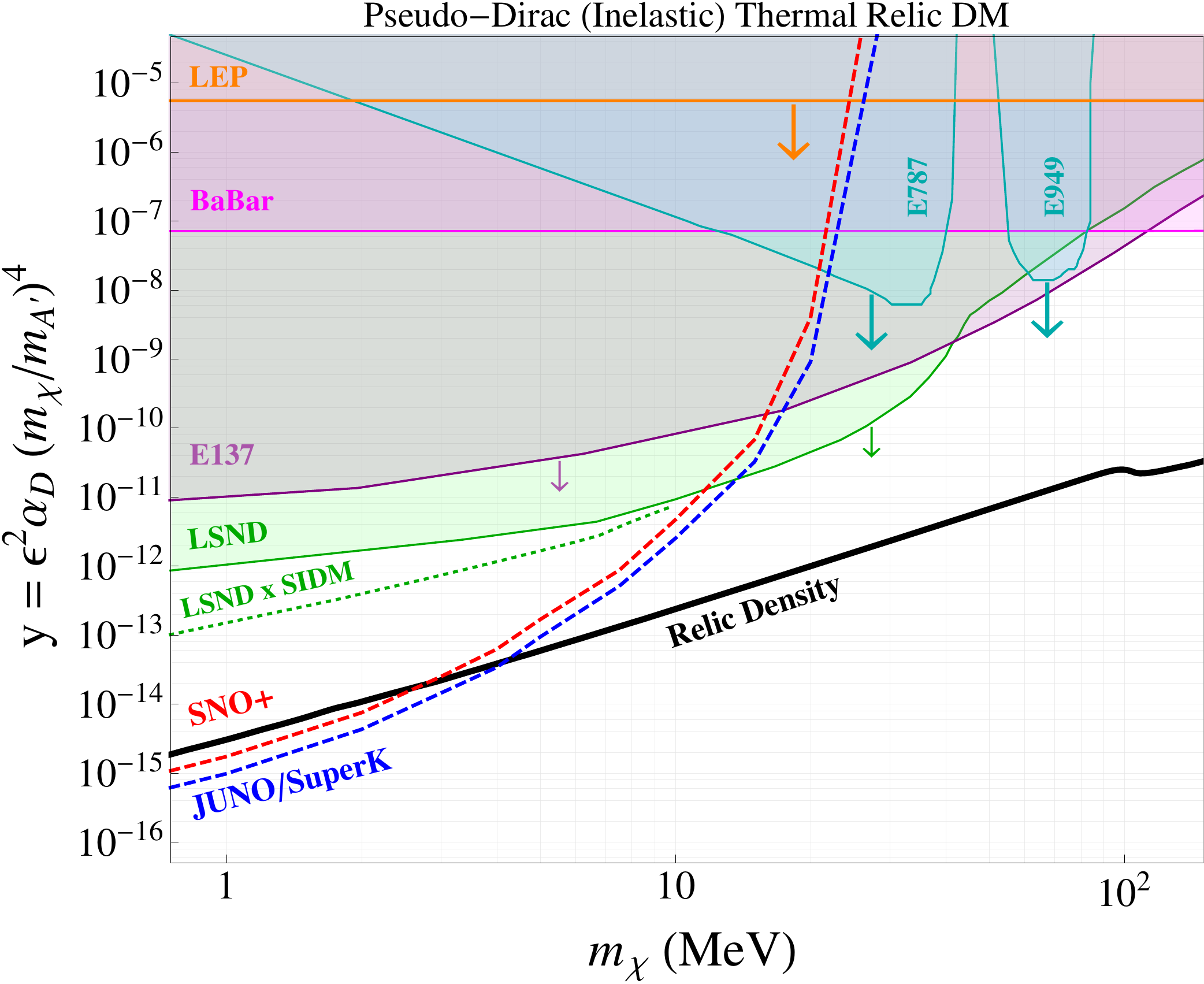}  ~~~
\includegraphics[width=8.6cm]{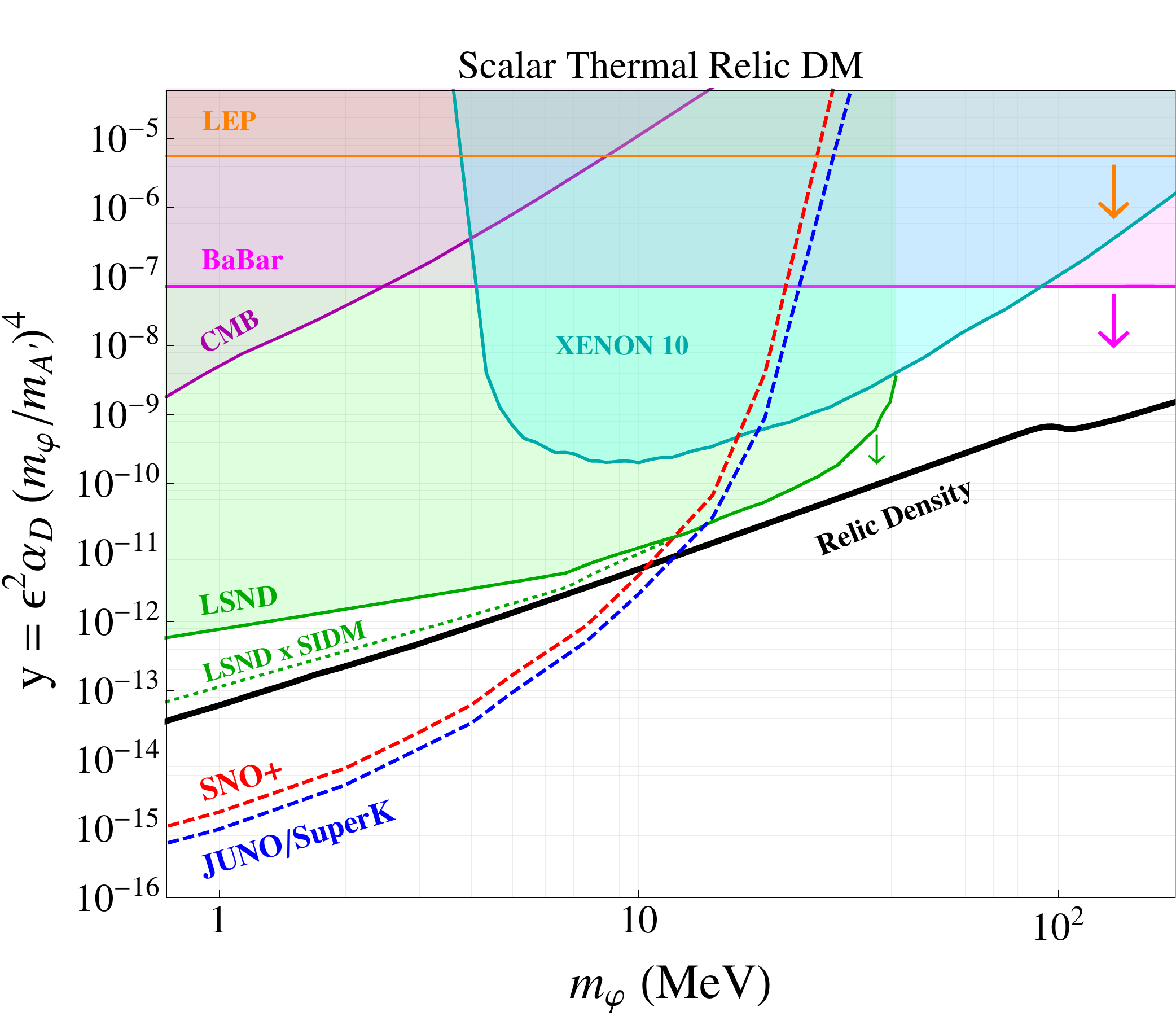}
  \caption{
 Sensitivity production for $10^{24}$ electrons with  100-MeV energies impinging on an aluminum target positioned 10 m near the SNO+, JUNO, and SuperK detectors -- since the latter two have comparable fiducial volumes, their projections are presented as a common curve. We conservatively assume thresholds of $E_R >10$ MeV for which the backgrounds are negligible. The CMB and direct detection constraints assume $\chi/\varphi$ constitutes all of the dark matter and regions above the relic curve correspond to parameter space for which each scenario can accommodate a subdominant fraction of the total DM (note that for subdominant DM, the CMB and direct detection bounds would also weaken). For the pseudo-Dirac scenario the relic curve was computed assuming only a small mass-splitting ($ 100~\keV <\Delta < m_{\chi/\varphi} $) between the states that couple to the $\apr$ so standard freeze out is largely unaffected, but scattering at direct detection
 experiments is kinematically inaccessible. Since Kaon, mono-photon, and beam-dump constraints don't scale as $y$, we conservatively adopt  $\alpha_D = 0.5$  and $m_{\chi/\varphi}/m_{\apr} = 3$ to not overstate these bounds; the plotted arrows show how the constraint moves when the product $\alpha_D (m_\chi/m_{\apr})^4$ is reduced by a factor of ten.  The dotted LSND $\times$ SIDM curve
 denotes where the LSND bound shifts if $\alpha_D$ is chosen to satisfy the bound on
 DM self interactions $\sigma_{\rm self}/m_\chi \lsim0.1~\cm^2/\gram$ instead of the nominal $\alpha_D = 0.5$ which is conservative
 in other regions of parameter space. Note that for scalar inelastic DM, the key difference relative to the 
 right panel is that the Xenon10 region disappears as the scattering can be turned off. 
  }
\label{fig:yplot}
\end{figure*}

\begin{figure}[t!]
 \vspace{0.cm}
 \hspace{-0.3cm}
\includegraphics[width=8.6cm]{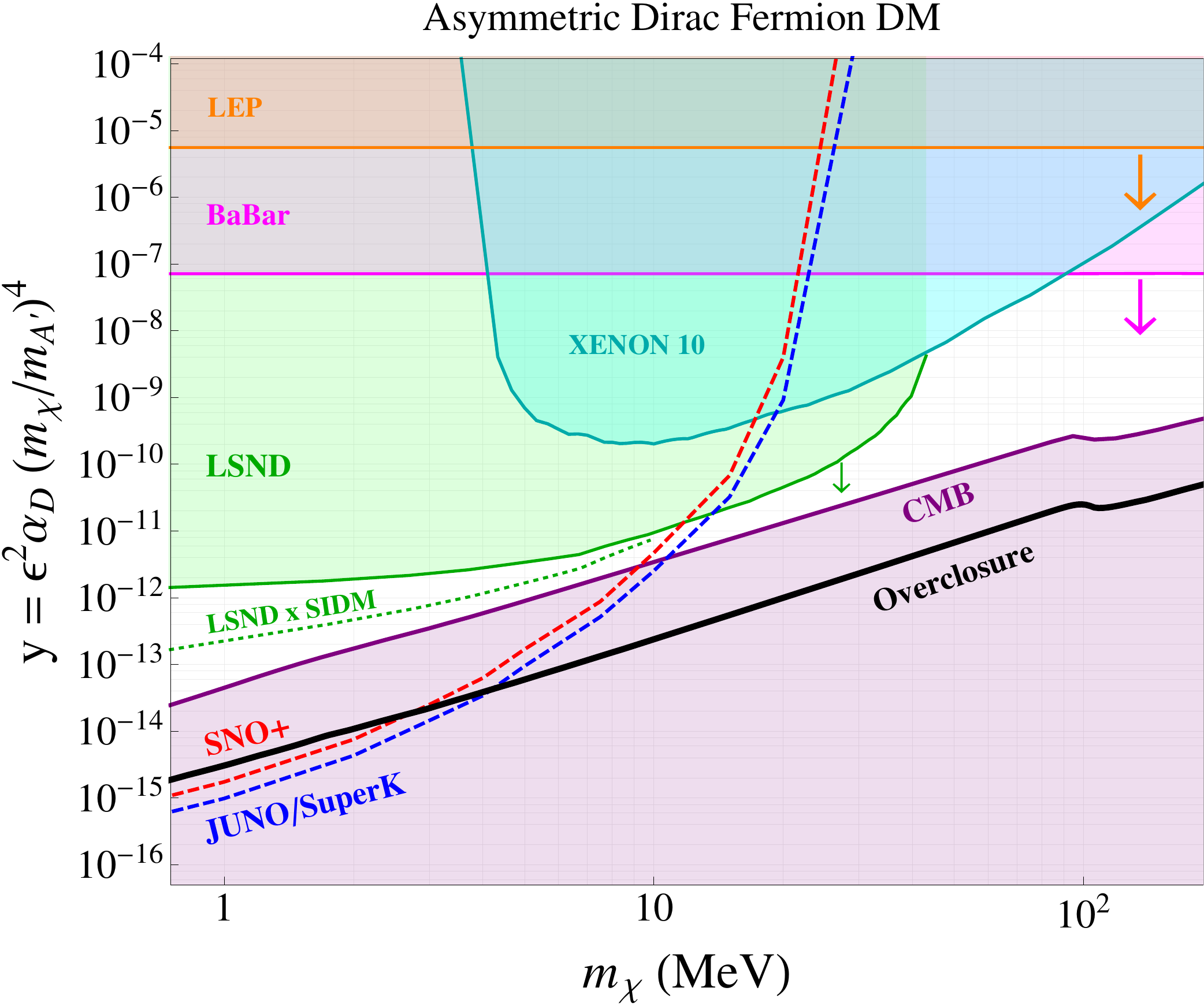}  ~~~
  \caption{
Same parametrization and conventions as Fig. \ref{fig:yplot}, but for asymmetric Dirac fermion DM; the symmetric version of this scenario
is ruled out by the CMB. Note that the relic line has been replaced with the {\it lower} bound from the CMB. Since the residual antiparticle density in the early universe scales as
 $n_{\bar \chi} \sim \exp(-\sigma v)$, a larger annihilation rate exponentially weakens
  the CMB constraint. Unlike Fig. \ref{fig:yplot}, all unshaded points are consistent with achieving the full DM abundance. For scalar asymmetric
 DM (not shown), the plot is similar, only the CMB bound is weaker and the ``overclosure" curve is shifted upwards to 
 match the ``relic density" curve in Fig.~\ref{fig:yplot} (right).
  }
\label{fig:yplot2}
\end{figure}


In this paper we propose a powerful new setup depicted schematically in Fig.~\ref{fig:schematic} --- the combination of a large underground detector such as those housed in neutrino facilities and a low-energy (10-100 MeV) but high intensity electron-beam --- which is capable of sharply testing the thermal origin scenario below $\sim$ few 10s of MeV in DM mass. While our proposal requires a substantial investment in an accelerator and beam-dump deep underground, it can significantly surpass the sensitivity of all other existing efforts in this mass range. This concept complements the DAE$\delta$ALUS light-DM  proposal \cite{Kahn:2014sra} involving an underground proton beam as well as other underground accelerator concepts \cite{Conrad:2013sqa,Aberle:2013ssa,Izaguirre:2014cza} with different physics goals.

For concreteness, we consider light DM that interacts 
with the visible sector through a kinetically-mixed \cite{Holdom:1985ag} massive dark-photon $A'$ \cite{Okun:1982xi}. Since
light DM requires a comparably light mediator to avoid overclosure, this starting point loses no essential generality and our results
are qualitatively similar for different mediators.
The most general renormalizable Lagrangian for this dark sector contains 
\be
 \label{eq:lagrangian}
{\cal L}_{\rm D} \supset 
 \frac{\epsilon_Y}{2} F^\prime_{\mu\nu} B_{\mu \nu} + \frac{m^2_{A^\prime}}{2} A^{\prime}_\mu A^{\prime\, \mu}+  {\cal L}_{ DM} ,
 \ee
where $A^\prime$  is the dark photon that mediates an abelian $U(1)_D$ force,  
 $F^\prime_{\mu\nu} = \partial_{[\mu,} A^\prime_{\nu]}$ 
 and  $B_{\mu\nu} = \partial_{[\mu,} B_{\nu]}$ 
 are  the dark and hypercharge field strength tensors, and $m_{\chi, {A^\prime}}$ are the appropriate dark sector masses. 
 After electroweak symmetry breaking, diagonalizing the gauge boson mass matrices induces a kinetic mixing with the
 photon field strength $\epsilon \equiv \epsilon_Y \cos\theta_W$, where $\theta_W$
 is the weak mixing angle.
The DM Lagrangian contains a fermionic or bosonic MeV-scale DM particle charged under  $U(1)_D$,
\be
{\cal L}_{ DM}=\left\{\begin{array}{c}\bar \chi ( i \displaystyle{\not}{D}- m_\chi) \chi, ~~~{\rm fermionic ~DM},  \\
|D_\mu \varphi |^2 - m_\varphi^2 \varphi^*\varphi,  ~~~{\rm bosonic ~DM,}\end{array}\right.
\ee
where $D_\mu = \partial_\mu + i g' A'_\mu$ is the covariant derivative. 
These simplest realizations of ${\cal L}_{D}$ assume a Dirac fermion or complex scalar DM states, 
but the model can readily be generalized to the ``split" states of Majorana/pseudo-Dirac fermions or real scalars, 
in which case $\apr$ can couple off-diagonally (inelastically) to the different mass-eigenstates and
 $m_{\chi(\varphi)}$ should be understood as a matrix acting on the split states. Moreover, each variation above
 can be particle/antiparticle asymmetric, which allows for weaker bounds from late-time annihilations into the CMB than the symmetric case \cite{Lin:2011gj}.

One of the most important questions for such a model is the hierarchy of masses in ${\cal L}_D$. If $m_{\apr} < m_{\chi/\varphi}$, the mediator is the lightest state in the dark sector, so it will decay into SM particles. In this regime, the annihilation process that
sets the relic density is $t$-channel (e.g. $\chi \bar\chi \to \apr \apr$) and, thus, independent of the mediator's coupling to the SM. 
However, if $m_{A'} > m_{\chi(\varphi)}$, then the relic density is achieved through $\chi \bar \chi \to e^+e^-$ annihilation, 
which proceeds via a virtual s-channel $A'$ exchange and depends on both DM and SM couplings to the mediator\footnote{In a certain
 region of parameter space, the $m_{\apr} > m_\chi$ scenario can still  achieve the observed relic abundance through $\chi \bar \chi \to \apr \apr$ annihilation in the ``forbidden" channel \cite{D'Agnolo:2015koa}, but 
this possibility is beyond the scope of our work. }. This latter scenario is predictive: since dark sector couplings are bounded by perturbativity, sufficient experimentally sensitivity to the SM-mediator coupling can discoverer or decisively rule out 
this class of direct-annihilation models.  

The leading $s$-wave annihilation cross section for Dirac DM is given by
\be
\sigma v = \frac{16 \pi \alpha\alpha_D m_\chi^2(1 + m_e^2/2m_\chi^2) }{ (m_{A'}^2 - 4 m_\chi^2)^2}  \sqrt{1 - m_e^2 / m_\chi^2},
\ee
while for the scalar case the annihilation is $p$-wave   
\be
\sigma v &=&    \frac{8\pi}{3} \frac{ \epsilon^2 \alpha \alpha_D m_\varphi^2 v^2 (1 + m_e^2/2m_\varphi^2)}{(m_{A'}^2 - 4 m_\varphi^2)^2   }  \sqrt{1 - m_e^2 / m_\varphi^2}~,~~  \label{eq:scalar-annihilation-rate-A}
\ee
and, hence, suppressed at late times (e.g. during CMB last scattering). 
Here $\alpha_D = (g')^2/(4\pi)$ and $v$ is the relative velocity between annihilating particles. 
These expressions are approximately valid for the inelastic variations of each scenario in the limit
where the mass splitting between annihilation partners is small compared to their mass. 
Equating this annihilation rate to $\sim 1~{\rm pbn}\times c$ gives 
an important relation among the parameters of the model that ensures the correct cosmological abundance. 

This class of models can easily be adapted to cover a wide range of variations that 
can arise in a generic dark sector with a thermal history. Beyond the spin of 
the DM and mediator, we can classify the dominant dark species by its present
day abundance (particle-antiparticle symmetric vs. asymmetric) and the nature of its
coupling to the mediator (elastic vs. inelastic). Since the experimental setup we advocate in this paper  
(direct accelerator production and downstream scattering) is largely insensitive to
 these differences, we regard our approach as a convenient simplified model; the sensitivity of 
  this approach can be mapped onto most thermal DM scenarios in the low mass range. 

In comparing these scenarios against the thermal relic target,
 it is useful to adopt the ansatz introduced in \cite{Izaguirre:2015yja} and define the dimensionless
 interaction strength  
\be
y \equiv  \epsilon^2 \alpha_D\left( \frac{m_{\chi/\varphi}}{m_{\apr}}\right)^4,~
\ee
which, up to small additive corrections, is proportional to the annihilation rate and insensitive to 
assumptions about ratios of individual Largangian parameters. Thus, plotting $y$ vs. $m_{\chi/\varphi}$ invariantly presents the thermal reach of various experimental bounds. Some experiments, like direct
detection, allow for unambiguous comparison with the thermal target since the DM-SM 
non relativistic scattering cross section is typically proportional to $y$. However, other constraints only
depend only on a subset of the parameters that constitute $y$ (e.g. collider production) so it's necessary to make conservative assumptions about the other parameters for a robust comparison with the 
thermal target -- for a discussion see Sec.~\ref{sec:constraints}.

For any  DM search involving accelerator production and downstream detection, the signal yield
 is proportional to the number of DM particles produced and 
the combined probability that the DM reaches the detector and scatters off a stationary target inside. 
If the mediator ($A'$) can kinematically decay to DM then the signal yield scales as 
\be
{\rm Signal} ~\propto {\rm POT} \times(P_{\rm ~Geom.\,Acc.})  \times  (N_{\rm Atoms~in ~det.}),
\ee
where POT stands for the number of particles-on-target, $ P_{\rm ~Geom.\,Acc.}$ is the probability that the produced DM path 
crosses the geometry of the detector, and $N_{\rm Atoms~in ~det.}$ is the number of target particles inside such a detector. 
The ideal light DM beam-dump experiment seeks to maximize the product of these factors, while minimizing background rates. 

Our proposal, which pairs a high-intensity low-energy electron-beam accelerator with a large underground detector,
Fig. \ref{fig:schematic}, has several key features that maximize signal and reduce potential  backgrounds:

\begin{enumerate}

\item The next generation of deep underground neutrino detectors are among the largest detectors ever built or proposed. 

\item Underground facilities, typically situated inside a mine at a few water-equivalent km underground, boast enormous overburdens, which exponentially suppress 
cosmic and other environmental backgrounds.

\item An electron beam can be advantageous over a proton beam for probing MeV-scale New Physics (NP) since beam-related backgrounds are significantly suppressed compared to the latter, as the production of neutrinos and neutrons is significantly less in the electron beam case.

\item For DM masses small compared to the beam energy, radiative production in electron-nucleus collisions is extremely forward-peaked and offers
excellent (order-one) geometric acceptance for fixed-targets placed in the vicinity of large neutrino detectors. 

\item A high-intensity electron-beam --- e.g. a continuous wave (CW) beam at existing facilities --- capable of running at mA currents combined with a large neutrino detector offers a significant improvement in luminosity compared to previous experiments.

\item Finally, the proposed light DM search scheme does not interfere with other physics goals of large underground detectors.
Moreover, for some facilities, a project of this scope and ambition may offer an entirely novel physics program -- 
especially for detectors whose main physics goals have already been met (as is the case for most solar neutrino detectors).

\end{enumerate}
  
For the remainder of this paper we narrow our scope to simple models in  
Eq.~(\ref{eq:lagrangian}), in which DM communicates with the SM through the vector portal. 
In Sec.~\ref{sec:sigprod} we discuss the production and detection of a lightly coupled new 
particle at underground facilities; Sec.~\ref{sec:facilities} discusses the potential promising facilities for realizing the proposal in this paper;
in Sec.~\ref{sec:backgrounds} we discuss the main backgrounds for such an experimental setup;
In Sec.~\ref{sec:constraints} we discuss existing experimental constraints on this parameter space and  the 
projected sensitivity of our setup for various detectors; finally in Sec.~\ref{sec:discussion} we offer some 
concluding remarks.
  



\section{Production and Detection}
\label{sec:sigprod}
\subsection{Invisibly Decaying Mediator}

In this section we follow a well-established routine for calculating the elastic scattering of light DM $\chi$ 
produced from the decay of on-shell mediators $A'$ \cite{Izaguirre:2013uxa, Izaguirre:2014dua}. The production of 
mediators occurs via radiative emission from electron-nucleus scattering,  $Z + e^- \to Z +e^- + A'$, and is calculable 
with a minimum of uncertainty, by standard QED methods. For the results in Sec.~\ref{sec:constraints} we calculate the production of DM at a fixed target using a modified version of \texttt{Madgraph} \cite{Alwall:2014hca} from Ref.~\cite{Essig:2010xa},
which includes electron-nucleus scattering with form-factors obtained from Ref.~\cite{Tsai:1973py}.

\begin{figure}[t!]
\includegraphics[width=8.5cm]{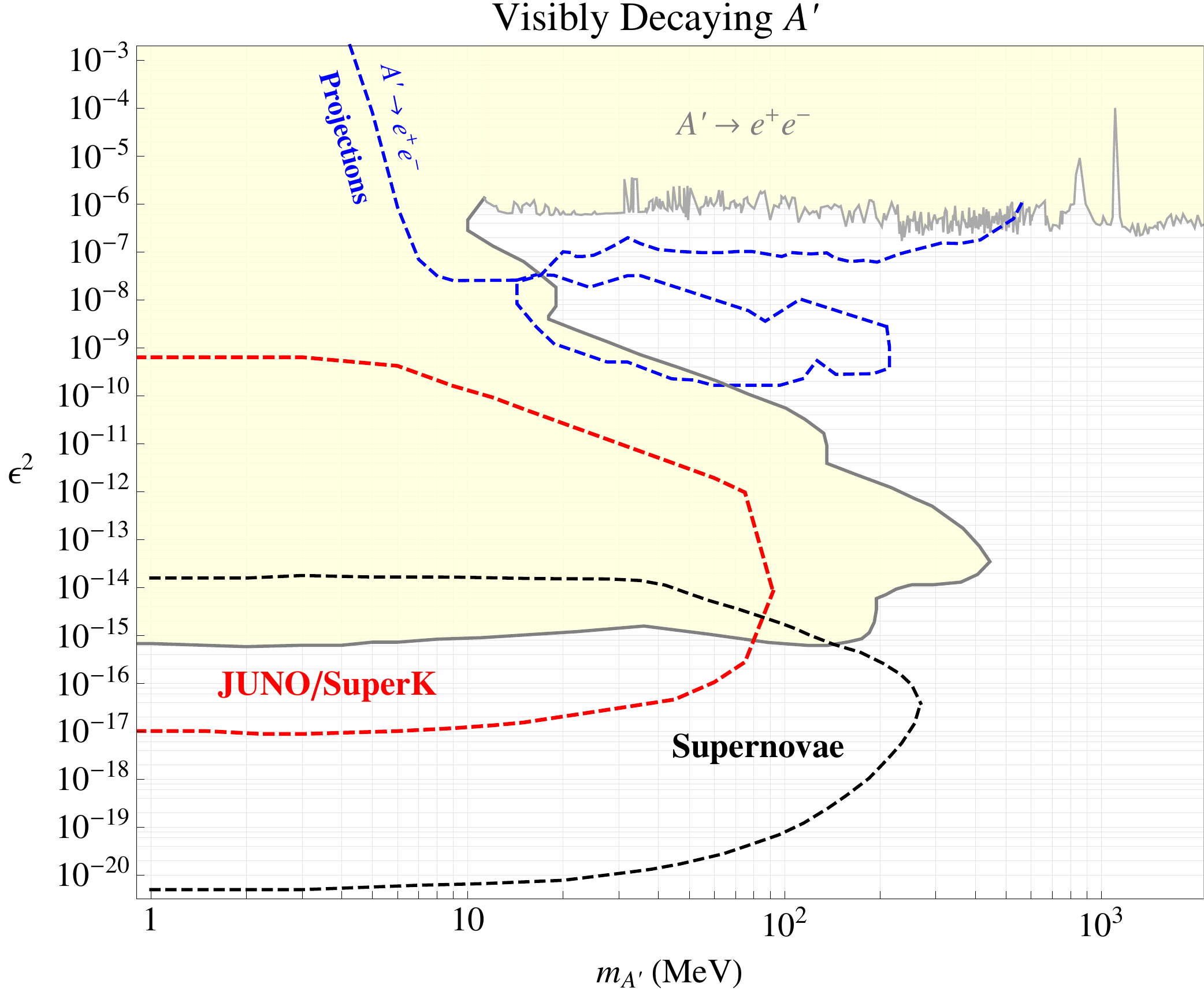} 
\caption{Constraints and projections for the visibly decaying $\apr \to e^+e^-$ scenario. The shaded region
represent existing bounds from beam dump and accelerator searches taken from \cite{Essig:2013lka}.  The red curve 
is  the 10 event yield projection for the same JUNO or SuperK setup described above, but deriving a signal yield from both visible decays and $\apr$ absorption. Comparable projections are obtained for other detectors considered above. The dashed blue curve represents projections for future visibly-decaying $\apr$ searches 
reviewed in \cite{Essig:2013lka} and the dashed black region represents model-dependent interpretations of supernova observations. If additional interactions allow the $\apr$ to thermalize with (heavier) DM states also produced in during the explosion, the energy loss bounds can be alleviated-- see discussion in Sec.~\ref{sec:constraints}. }
\label{fig:visplot}
\end{figure}

The prompt decay of $A'$ mediators to light DM, $A'\to \chi\bar\chi$, creates a flux of (semi-)relativistic $\chi$ 
particles that subtends the detector volume. 
For invisibly decaying mediators, the signal consists of detector particle recoils induced by $\apr$ daughter particles -- e.g. $\chi e \to \chi e$ mediated by virtual $\apr$ exchange. The scattering yield for this process is 
\be \label{eq:scatyield}
~~~ Y_{S}  =  \frac{N_{\apr} n_{e} \varepsilon }{N_{s}}\,
\sum_{i =1}^{N_{s}} 
   \int_{E_{c}}^{E_{0}} \!\!\! \!   dE\, \frac{d\sigma_S}{dE_R}(E_i) \ell_i  ~  \Theta( \theta_D - \theta_i)~,~
\ee
where $E_0$ is the $e^-$ beam energy, $E_c$ is the recoil ``cut" threshold, $N_{\apr}$ is the total number of $\apr$ produced, $E_{i}$ is the daughter DM 
energy, $E_R$ is the target's recoil energy, $n_e$ is the target (electron) number density, and $\ell_i$ is the DM
path length through the detector. For our signals of interest, typically $E_R \gsim 10$ MeV,  the detection efficiency $\varepsilon$ is of order-one. The theta function enforces geometric 
acceptance by admitting only particles whose angle $\theta_i$ with respect to the beam line falls 
inside $\theta_D$, the angle subtended by the detector. We have presented this rate as an
average over $N_s$  simulated monte carlo events of $\apr$ production and decay in the target.  
The nature of the scattering cross section can vary depending on energy and detection methods. 
For most large neutrino detectors, the dominant signal is from electron recoils. Assuming Dirac fermion
DM,  the recoil spectrum is 
\be
\frac{d \sigma_S}{dE_R}= \frac{\alpha_D \epsilon^2}{\alpha} \times
\frac{8\pi\alpha^2m_e(1-E_R/E_i)}
{(m^2_{A'} + 2m_eE_R )^2}~,
\ee
in the $E_i \gg m_\chi$ limit; a parametrically similar expression can be obtained for inelastic or scalar DM variations \cite{Izaguirre:2014dua}.
Other signatures may include elastic and quasi-elastic scattering of $\chi$ on nucleons, 
and nuclear elastic recoils.  For the inelastic-DM scenario, the process $\chi_1 e \to \chi_2 e$
is well-approximated by the above formula in the limit where the mass splitting is small compared  
to the incident DM energy.

\subsection{Visibly Decaying Mediator}
\noindent For the visibly decaying mediator  ($m_{A'} < 2 m_\chi$),  
the signal consists of both decays and absorptions. 
For a simulated population of $N_{\apr}$ mediator particles, the total yield is 
$Y = Y_{A} + Y_{D}$, where the effective absorption ($A$) and decay ($D$) contributions are 
\be
&& \hspace{-0.3cm}  Y_{A}  =  \frac{ N_{\apr } n_{e} \varepsilon }{N_{s}}\,\sum_{i =1}^{N_{s}} e^{-\frac{\Gamma d_i   }{c \gamma_i}}  \!   \int_{E_{c}}^{E_{0}} \!\!  dE\, \frac{d\sigma_A}{dE}(E_i) \ell_i  \Theta( \theta_D -\theta_i)
  \nonumber  \\
&& \hspace{-0.3cm} Y_{D}  =  \frac{N_{\apr} \varepsilon}{N_{s}}\,\sum_{i =1}^{N_{s}} e^{-\frac{\Gamma d_i }{c \gamma_i}} \left(1-  e^{- \frac{   \Gamma \ell_i }{c \gamma_i}} \right) \Theta( \theta_D -\theta_i)~.
\ee
Unlike in Eq.~(\ref{eq:scatyield}), here $E_i/\theta_i$ are now the $\apr$ energy/angle of the $i^{\rm th}$ $\apr$ produced in the simulation, $\gamma_i = (1-v^2)^{-1/2}$ is the $\apr$ boost factor, 
 $d_i$ is the path traveled between target and detector, and the visible $\apr$ width is 
\be
\Gamma = 
\frac{    \epsilon^2 \alpha    \, m_{\apr}  }{3 \pi }   (1 + 2 m_e^2/m_{\apr}^2   )
  \sqrt{1 - 4 m_e^2 / m_{\apr}^2 } ~.
\ee
The absorption cross section for the $\apr e \to \gamma e$ 
process with a stationary detector electron is 
\be
\frac{d\sigma_A}{dE_R} = \frac{ m_e \langle |{\cal A}|^2\rangle }{ 32 \pi |p^*|^2 s}~,~
\ee
where $E_R$ is the electron recoil energy, 
$s = (m_e + E_i)^2$,   $|p^*|^2 = [  ( s - m_e^2 - m_{\apr}^2)^2 - 4 m_e^2 m_{\apr}^2 ] / 4s$ is the CM three-momentum
in the lab frame, and the absorption amplitude ${\cal A} = {\cal A}_s + {\cal A}_t$ is a sum of  $s$ and $t$ channel
amplitudes 
\be
{\cal A}_t \!\! &=&  \!\! \frac{\epsilon e^2 }{(t - m^2_e)} \bar u(p_R)\gamma^\mu (\displaystyle{\not}{p}_T - \displaystyle{\not}{p}_\gamma  + m_e) \gamma^\nu u(p_T)  \varepsilon^{\apr}_\mu \varepsilon^\gamma_\nu,~~~~~~\\ 
{\cal A}_s \!\! &=& \!\!\frac{\epsilon e^2 }{(s - m^2_e)} \bar u(p_R)\gamma^\mu (\displaystyle{\not}{p}_T + \displaystyle{\not}{p}_{\apr}  + m_e) \gamma^\nu u(p_T)  \varepsilon^{\apr}_\nu \varepsilon^\gamma_\mu,~~~~~~
\ee
where $p_{T/R}$ are the four-momenta of the target/recoil $e^-$, $p_{\apr}$ is the $\apr$ four-momentum, and $\varepsilon^{\apr,\gamma}$ are gauge boson
polarization vectors.




\section{Suitable Facilities}
\label{sec:facilities}
%
%
%

The experimental strategy that we propose relies on production of light DM by an electron beam impinging on a target or beam dump. Further detection downstream is realized at a large detector. The advantage of this approach is that it permits use of an existing or planned detector. We now discuss existing and upcoming detecting facilities which could be suitable to meet the physics targets outlined in this paper.

\subsection{Large water-based \v Cerenkov detectors}

\noindent{\bf \em Super-Kamiokande}

\noindent The Super-Kamiokande (SuperK) detector \cite{Fukuda:2002uc} in the Kamioka mine in Japan 
contains a 50,000-ton water \v Cerenkov detector, at a depth of 2700 m.w.e.. It was designed to measure neutrino rates from different sources such as those from astrophysics and the atmosphere; and for proton decay. It consists of a welded stainless-steel tank 39 $\rm{m}$ in diameter, it stands 42 $\rm{m}$ tall, and 
is sensitive to electron recoils above 4.5 MeV. An important feature of this type of detector
is directional sensitivity, which can be used to further suppress environmental backgrounds. 
Super-K is the largest among the existing detectors suitable for our proposal. It currently hosts 
a very low-power medical-type electron linac for calibration purposes \cite{Nakahata:1998pz}. 
 \\
 \\
 
\noindent {\bf\em Hyper-Kamiokande}

\noindent The next large water  \v Cerenkov detector at Kamioka mine is the Hyper Kamiokande project \cite{Abe:2011ts}, intended to also measure neutrino rates, look for proton decay, in addition to being the far detector for a long baseline neutrino oscillation experiment using neutrinos originated from the upgraded J-PARC. Its design is based on the same detection technology that SuperK featured. It has a $40 \rm{m}\times 54 \rm{m}\times 250 \rm{m}$ detector volume, with a fiducial volume $\sim 20$ times greater than that of SuperK, located at a depth of 1750 m.w.e..
 
\subsection{Large liquid scintillator detectors}
\noindent {\bf \em Borexino}  

\noindent The borexino detector is located in the Gran Sasso laboratory in Italy, under 3800 m.w.e. and contains a large volume 
liquid scintillator in a 8.5 m diameter vessel \cite{Alimonti:2008gc}. The detector registers energy depositions (mainly $\gamma$-events and electron recoils, and with some quenching factor penalty it can also 
observe proton recoils) and offers good energy resolution with a low threshold ($\sim$200 keV), which allowed Borexino to  measure 
different components in the solar neutrino flux. It is currently being considered as a short-baseline detector for neutrinos emitted
from an intense radioactive source \cite{Borexino:2013xxa}.  
\\ 

\noindent {\bf \em Kamland} \\
\noindent  The Kamland detector is located in Kamioka mine, a 13 m sphere filled with liquid scintillator. 
It has played an important role in the precision determination of neutrino mass splitting and mixing \cite{Eguchi:2002dm}. 
It currently has been re-profiled as a tool to search for the neutrino-less double-beta decay \cite{Gando:2012zm}. 
\\ 

\noindent {\bf \em SNO+}
 
\noindent SNO+, the upgraded Sudbury Neutrino Observatory (SNO) is a 12 $\rm{m}$ diameter liquid scintillator, situated at $\approx$ 6000 m.w.e. \cite{Chen:2008un}. The detector features a significant light-yield upgrade over its predecessor, and chief amongst its current science goals are  neutrino-flux measurements from the $pp$, $CNO$ cycle, from supernovae, in addition to searching for neutrino-less double-beta decay. SNO+ energy thresholds are expected to be near 200 keV.
\\

\noindent {\bf \em JUNO} 

\noindent The Jiangmen Underground Neutrino Observatory (JUNO) is a multi-purpose neutrino oscillation experiment, \cite{He:2014zwa} situated 700 $\rm{m}$ underground and  currently under construction, with data-taking expected to commence in 2020. The detector itself is a 35.5 $\rm{m}$ diameter liquid scintillator embedded in a water pool which will also serve as a \v Cerenkov muon veto.
\\
\medskip

\noindent   {\bf \em DUNE} 

\noindent Another interesting opportunity may arise from the DUNE(LBNF) project. 
A detector containing 40 kton of liquid argon is planned to be built at the Sanford Underground Research Facility (SURF), 
at a depth of  4300 m w.e.  While the main goal of this detector will be observing long baseline neutrino oscillations, 
it can also be used in conjunction with a linac at SURF.

\subsection{Large-scale dark matter detectors and $0\nu2\beta$ detectors}

All of the detectors described above have over $100$ tons (sometimes several kilotons) of detecting material. It is also 
worth noting that underground facilities often house very large, $O(1~{\rm ton})$, detectors aimed at the direct dark matter 
detection and/or neutrino-less double-beta decay processes. These include XENON1T \cite{Aprile:2012zx}, 
LUX \cite{Akerib:2013tjd}, DEAP-3600 \cite{Amaudruz:2014nsa}, future LZ experiment \cite{Malling:2011va} --
all devoted to DM direct-detection for $> 10$ GeV masses -- as well as other experiments such as 
{\em e.g.}  EXO \cite{Auger:2012ar} and CUORE \cite{Artusa:2014lgv} that take aim at $0\nu2\beta$. Although the active 
mass in such detectors is smaller than neutrino detectors, some of them have remarkably low energy thresholds, and are capable of detecting 
coherent scattering on nuclei. This coherence typically adds another enhancement 
factor of $Z\sim O(50)$, compared to incoherent scattering on electrons, which helps to offset their small mass. 
In many of the underground facilities these detectors co-exist with large neutrino detectors, and therefore it may be possible to
perform an accelerator-based search for light DM using multiple detectors hosted in the same laboratory.



\section{Backgrounds for Underground Accelerator Beam Dump searches}
\label{sec:backgrounds}

One of the advantages of the experimental setup proposed in Sec.~\ref{sec:sigprod} is the clean environment that the underground facilities offer. In this section we discuss the relevant backgrounds that give rise to SM muons and neutrinos, which can mimic the signal from light DM scattering off electrons. We split the discussion into beam-related backgrounds arising from reactions in the target or dump and environmental backgrounds arising from ambient radioactivity and cosmic-rays traversing the detector. 

We first discuss beam-related backgrounds. For the proposal in this study which relies on an electron beam running at an energy below $m_\pi$, beam-related backgrounds may arise from a few sources. Direct production of neutrinos in reactions due  to weak neutral or charged 
currents, $e N \rightarrow e N \bar\nu \nu$ and  $eN\rightarrow \nu_e N' + X$, followed by neutrino scattering in a detector, $\nu N\rightarrow \nu N'+X$
or $\nu e \to \nu e$, is small. Indeed, the production cross section is $\approx fb/\rm{nucleon}$, \cite{Agashe:2014kda}, giving only up to  $10^{5}-10^{6}$ neutrinos with $E_\nu \sim $~beam energy,  produced in one year's running and assuming an aluminum beam-dump. For a neutrino to fake the DM signal, it needs to subsequently scatter in the detector situated behind the dump. The probability for a neutrino to scatter off an electron in the detector is given by $\sigma_{\nu e\rightarrow \nu e} n_{D}\ell_D$, where $n_D$ and $\ell_D$ are the number density and longitudinal length of the detector. The scattering cross section scales as $G_{F}^2 E_\nu m_e$. Even if the neutrino produced in the target carries most of the beam energy, the event rate from this type of background will be well below the atmospheric neutrino background. 

A more serious source of beam-related backgrounds is the spallation of nucleons by the electron beam. This may lead to a 
neutron scattering background and to $G_F$-unsuppressed production of neutrinos by resulting $\beta^{(+)}$ emitters. 
The inelastic cross sections of electrons on nuclei can reach $\sim mb$ range, and therefore, the number of produced 
neutrons and neutrinos will significantly exceed the direct production discussed in the previous paragraph. 
However, extra neutrons produced this way will be intercepted by some meters of rock separating the production point 
from the detector. The neutrino energies will be limited to $E_\nu \lsim 5$ MeV, typical for $\beta$-processes, and can be 
removed by a simple recoil energy cut. Therefore, both types of backgrounds are reducible. 
Moreover, a judicial choice of beam dump material can reduce this contamination even further, so
 we conclude that  the beam-related backgrounds for the experimental setup we advocate for are negligible.

The nature of the accelerator beam plays a crucial role in whether environmental backgrounds are important or not. The higest intensity beams may only be attained by using a continuous wave (CW) beam. In that scenario, given the $\mathcal{O}(1)$ duty cycles common to a CW beam, using timing to reduce cosmic-related backgrounds is not an available handle, unlike at a pulsed beam. Nevertheless, environmental backgrounds can still be reduced to a negligible level through a combination of existing techniques. We base the following feasibility argument  on the lessons learned from the results by the Borexino collaboration. First, the radiated massive $A'$ from the reaction $eN\rightarrow e N A'$ will carry most of the beam energy. 
Assuming a 100 MeV electron beam, this means that one could use a high electron recoil-energy threshold in the detector. In particular, a $E_R > 5 \MeV$ threshold would eliminate all radiogenic backgrounds that Borexino has encountered thus far. This still leaves cosmic ray muons, cosmogenic neutrons, and solar neutrinos originating from the  $^{8}$B$\rightarrow ^{8}$Be$ + e^+ + \nu_e$ reaction in the sun as a potential source of contamination for a light DM scattering signal. However, these could be significantly reduced through a combination of shielding, fiducialization, timing veto following activity in the detector (for stopped muons), and directionality, as was attained in Ref.~\cite{Bellini:2012kz} (see Fig. 5). 
In Sec.~\ref{sec:constraints} we conservatively adopt a  $10$~MeV energy threshold, and assume a background-free experimental setup for our sensitivity estimates.




\section{Projected Sensitivity and Results}
\label{sec:constraints}

We now turn our attention to the existing constraints on DM in the (1-100) MeV mass range for the class of models in Eq. (\ref{eq:lagrangian}). 
One of the strongest limits for DM masses below $m_\pi$ comes from a reinterpretation of results from surface-level neutrino experiments, in particular, LSND's measurement of neutrino-electron neutral current scattering \cite{Auerbach:2001wg} and recast by \cite{deNiverville:2011it}. Production of DM in such a setup proceeds from protons impinging on a fixed target. The latter gives rise to mesons in the final state, primarily $\pi^0$s which consequently can decay to either a DM pair or to a mediator via   $\pi^0\rightarrow \gamma \bar\chi \chi / \varphi \varphi^*$ or $\pi^0\rightarrow \gamma A'$, and finally with decay of the $A'$ into a DM pair. Once produced, DM detection proceeds analogously to neutrinos, with DM traveling through a near or far detector that is sensitive to the scattering process $\chi T \rightarrow \chi T$. Here $T$ could be a nucleus, nucleon, or an electron. The results above use the electron-scattering channel. A proposal for further running at MiniBoone could further improve the constraints at neutrino experiments for higher mass DM \cite{Dharmapalan:2012xp}.  

In the context of DM searches, one of the main limitations of neutrino factories is that neutrinos can be a {\it background} to the DM signal. In contrast, electron beams do not suffer from this limitation \cite{Izaguirre:2013uxa}. In fact, for an electron beam with energy of 100 MeV, neutrinos originating from muons or pions are no longer kinematically accessible. One of the strongest constraints on light DM comes in fact from E137 in a manner analogous to neutrino experiments \cite{Batell:2014mga} where DM production in the target proceeds via $e N \rightarrow e N A'/\bar\chi\chi/\varphi \varphi^*$. E137 sets a constraint on DM from the non-observation of anomalous electron-recoil events in a detector $\sim 300$ meters downstream of the target. 

Another relevant constraint to the parameter space that we focus on is that from B-factories. In particular, Babar's search for untagged $\Upsilon(3S)$ decays to $\gamma + invisible$ \cite{Aubert:2008as} was reinterpreted by Refs.~\cite{Izaguirre:2013uxa, Essig:2013vha} to place limits on DM where the DM could be produced in the reaction $e^+ e^- \rightarrow \gamma A'$ (and similarly for off-shell production of DM).
Kaon decays, and in particular the missing energy mode $K^+ \to \pi^+\nu\bar \nu$ also impose some consitraints on the parameter space of the model via $K^+ \to \pi^+\bar\chi\chi/\varphi \varphi^*$. 

While standard nuclear direct detection experiments are not sensitive to DM masses near the $\MeV$ scale, electron scattering does constrain masses this light. While these searches are background-limited, they still set a conservative limit. In particular, Ref.~\cite{Essig:2012yx} reinterprets XENON10 electron data to set strong limits.

Other constraints on the kinetic mixing scenario can be derived from LEP which accurately measured the $Z$ boson mass. If the kinetic mixing parameter
is too large it can induce an unobserved shift in $m_Z$, which constrains $\epsilon \lsim 10^{-2}$ \cite{Hook:2010tw}.

In Figs.~\ref{fig:yplot} and \ref{fig:yplot2}  we present the expected sensitivity of the combination of an 100 $\MeV$ electron beam impinging on a beam-dump situated $\approx 10$ meters from a large underground detector, such as that of the SNO+, JUNO, and Super-K experiments. These results are plotted as a function of the DM mass and the variable
$y = \alpha_D \epsilon^2 (m_\chi^4/m_{A'})^4$ which, up to small additive corrections, is proportional to the annihilation rate. With this anzatz, the thermal relic target is invariant on the $y$ vs. $m_\chi$ plane and 
we don't need to assume any particular values of individual inputs for a robust comparison with experiment. 

However, some constraints on $\epsilon$ (e.g. from LEP or kaon decays) don't depend on $\alpha_D$ or on the mass ratio $m_\chi/m_{\apr}$, we must make conservative assumptions for these inputs in order to construct $y$ for these bounds. For the predictive regime of interest, $m_{\apr} > 2 m_\chi$, the thermal
relic annihilation rate is proportional to $\epsilon^2 \alpha_D m_\chi^2/m_{\apr}^4$, so we  choose both $\alpha_D$ and the ratio $m_\chi/m_{\apr}$ to be ${\cal O}(1)$, which reveals the maximum parameter space on the $y$ vs. $m_\chi$ plane; it is always trivially easy to overstate these bounds by choosing smaller values of these quantities,
which shifts the $y$ contours for collider and kaon bounds downwards by a large amount, so this approach is the most conservative. Note that Bullet-cluster constraints on DM self-interactions \cite{MiraldaEscude:2000qt, Bionta:1987qt} set an upper bound on $\alpha_D$ and therefore on 
$y$, which we include in Fig. ~\ref{fig:yplot}. 
 
Figure~\ref{fig:yplot} considers two scenarios: scalar thermal relic DM (right) and Majorana-like (or pseudo-Dirac) DM (left) with a mass-splitting\footnote{For MeV scale DM, the splitting can be much smaller $\sim {\cal O}(\keV)$ to shut off direct detection bounds, but a larger splitting may be necessary in order to 
sufficiently depopulate the excited state before CMB freeze-out \cite{IDMcosmology}, so we conservatively adopt 100 keV.} $\Delta \gsim 100 ~ \keV$. The latter scenario is not subject to constraints from direct detection -- because the upscatter is kinematically forbidden at tree level -- or from
the CMB because the annihilation is exponentially suppressed when the heavier, unstable annihilation partner is depopulated before last-scattering. 
For scalar thermal relic DM,  the minimal symmetric realization is cosmologically viable because the $p$-wave annihilation rate is sharply redshifted at the epoch of recombination, thereby greatly alleviating the CMB bound. 

In Fig.~\ref{fig:yplot2} we show a similar plot for asymmetric Dirac fermion DM coupled to a dark photon. Unlike the particle-symmetric Dirac scenario (not shown), which is ruled out by the CMB for DM masses below 10 GeV \cite{Galli:2009zc,Galli:2011rz,Finkbeiner:2011dx,Madhavacheril:2013cna,Ade:2013zuv},  this variant remains viable because annihilations at early times exponentially deplete the antiparticle density before last scattering so the energy injected into the CMB at 
late times is reduced. Since the antiparticle density scales as $n_{\bar \chi} \sim \exp(-\sigma v) \propto  \exp(-y)$, larger values of $y$ are less constrained. so the CMB imposes a lower bound on $y$ for this class of models. In setting  this limit, we demand that $\sigma v /m_\chi \sim 10^{-28} \cm^3/\s/\GeV$ \cite{Madhavacheril:2013cna} while fixing the observed DM abundance at late times \cite{Lin:2011gj}.

The projected reach of the proposed setup is quite significant. For DM lighter than 10s of MeV, our proposed production and re-scattering search can exceed the LSND reach by many orders of magnitude and completely cover all viable thermal scenarios whose freeze-out abundance is set by 
$\chi \bar\chi (\varphi \varphi^*)\to e^+e^-$ annihilation. Only a dedicated missing momentum fixed-target scheme, Ref. \cite{Izaguirre:2014bca}, 
can offer more sensitivity since the yield for that setup scales as $ \epsilon^2$ instead of $\epsilon^4$.

Finally, for completeness we also include the constraints on the visibly-decaying $A'$ scenario in Fig. \ref{fig:visplot}. We observe that the proposed setup would be the most sensitive direct probe of this parameter space, greatly improving the reach of previous beam dump constraints. Although this region is naively excluded 
by theorists' interpretations of the supernova cooling bounds from SN-1987a, this constraint  
is model dependent, so we present it as a dashed black line in Fig. \ref{fig:visplot}.
For example, if in addition to $A'$ there is an additional mediator $A''$ and light DM $\chi$ that are fully thermalized during the SN explosion, the $A'$-$\chi$ interaction can also keep $A'$ thermalized and quench 
the energy loss through $A'$ emission. Therefore, the sensitivity reach to visible decays adds value to the proposed set-up.




\section{Discussion}
\label{sec:discussion}

This paper proposes a new experimental set-up to probe MeV-scale DM. The powerful combination of a low-energy high-intensity electron beam situated near a large scale detector underground can be complementary to existing proposals using electron direct detection \cite{Essig:2012yx, Hochberg:2015pha}, fixed-target missing momentum \cite{Izaguirre:2014bca} and future B-factories \cite{Essig:2013vha} (for similar proposals using proton beams underground see Ref~\cite{Izaguirre:2014cza} and Ref.~\cite{Kahn:2014sra}). In particular, it is capable of surpassing the sensitivity of LSND, a surface-level neutrino detector benefiting from a high-intensity proton beam, and decisively probing thermal DM lighter than $\sim$ 10\,s of MeV in some of the simplest light DM scenarios.  Given 
that many underground detectors have already achieved their original physics goals, a light-DM  search via accelerator production may add an exciting new application to these facilities.

{\em Acknowledgments.}
We thank Philip Schuster, and Natalia Toro for helpful conversations. 
The Perimeter Institute for Theoretical Physics is supported by the Government of Canada through Industry Canada
and by the Province of Ontario.

\bibliographystyle{apsrevM}
\bibliography{UndergroundLinac}

\end{document}